%% file: monoHbb_paper.tex
\newcommand*{\ATLASLATEXPATH}{latex/}
\documentclass[cernpreprint,texlive=2011,txfonts,UKenglish]{\ATLASLATEXPATH atlasdoc}
\pdfoutput=1
\usepackage[subfigure]{\ATLASLATEXPATH atlaspackage}
\usepackage{\ATLASLATEXPATH atlasbiblatex}
\usepackage{\ATLASLATEXPATH atlascontribute}
\usepackage{\ATLASLATEXPATH atlasphysics}
\graphicspath{{logos/}{figures/}}
\usepackage{monoHbb_paper-defs}
\usepackage{multirow}
\usepackage{epstopdf}
\usepackage{paralist}

\input{monoHbb_paper-metadata}

\hypersetup{pdftitle={Search for dark matter in association with a
    Higgs boson decaying to b-quarks in pp collisions at sqrt{s} = 13
    TeV with the ATLAS detector},pdfauthor={The ATLAS Collaboration}}

\begin{document}
\maketitle

\section{Introduction}
\label{sec:introduction}
\input{sec-introduction}

\section{ATLAS detector}
\label{sec:detector}
\input{sec-detector}

\section{Data and simulation samples}
\label{sec:data-mc}
\input{sec-data-mc}

\section{Object reconstruction}
\label{sec:object}
\input{sec-object}

\input{sec-analysis}
\input{sec-systematics}

\section{Results}
\label{sec:results}
\input{sec-results}

\FloatBarrier
\section{Conclusion}
\label{sec:conclusion}
\input{sec-conclusion}

\section*{Acknowledgments}
\input{Acknowledgements}

\printbibliography
\newpage
\input{atlas_authlist}

\end{document}

%% file: monoHbb_paper-metadata.tex
\AtlasTitle{Search for dark matter in association with a Higgs boson decaying to $b$-quarks in $pp$ collisions 
at $\sqrt{s} = 13$ TeV with the ATLAS detector}

\author{The ATLAS Collaboration}

\usepackage{authblk}

\AtlasRefCode{EXOT-2015-23}
\AtlasNote{ATLAS-EXOT-2016-023}
\PreprintIdNumber{CERN-EP-2016-181}

\AtlasJournal{Phys.\ Lett.\ B.}
\AtlasJournalRef{Phys.\ Lett.\ B. 765 (2017) 11}
\AtlasDOI{10.1016/j.physletb.2016.11.035}

\AtlasAbstract{%
A search for dark matter pair production in association with a Higgs
boson decaying to a pair of bottom quarks is presented, using 3.2
$fb^{-1}$ of $pp$ collisions at a centre-of-mass energy of 13 TeV
collected by the ATLAS detector at the LHC. The decay of the Higgs
boson is reconstructed as a high-momentum $b\bar{b}$ system with
either a pair of small-radius jets, or a single large-radius jet with
substructure. The observed data are found to be consistent with the
expected backgrounds.  Results are interpreted using a simplified
model with a $Z'$ gauge boson mediating the interaction between dark
matter and the Standard Model as well as a two-Higgs-doublet model
containing an additional $Z'$ boson which decays to a Standard Model
Higgs boson and a new pseudoscalar Higgs boson, the latter decaying
into a pair of dark matter particles.}

%% file: sec-introduction.tex
Although dark matter (DM) constitutes the dominant component of matter
in the universe, little is known about its properties and particle
content~\cite{Bertone:2004pz}. The leading hypothesis suggests that
most DM is in the form of stable, electrically neutral, massive
particles with cosmological constraints indicating that DM
interactions with Standard Model (SM) particles occur at a weak scale
or below~\cite{Steigman:1984ac}. Collider-based searches for the
particle content of DM provide important information complementary to
that from direct and indirect detection
experiments~\cite{Bauer:2013ihz}.

A traditional dark-matter signature at a proton--proton collider is
one where one or more SM particles, $X$, are produced and detected,
recoiling against missing transverse momentum -- with magnitude \MET~
-- associated with the non-interacting DM candidate.  A number of
searches at the Large Hadron Collider (LHC)~\cite{Evans:2008zzb} have
been performed recently, where $X$ is considered to be a hadronic
jet~\cite{Aad:2015zva, Khachatryan:2014rra}, $b-$ or $t-$
quarks~\cite{Aad:2014vea, Khachatryan:2014uma, Khachatryan:2015nua}, a
photon~\cite{Aad:2014tda,Chatrchyan:2012tea,Khachatryan:2014rwa,Khachatryan:2014tva},
or a $W/Z$
boson~\cite{ATLAS:2014wra,Aad:2014vka,Aad:2013oja,Khachatryan:2015bbl}. The
discovery of a Higgs boson, $h$~\cite{Aad:2012tfa,Chatrchyan:2012ufa},
provides a new opportunity to search for DM production via the
$h+\MET$ signature~\cite{Petrov:2013nia,Carpenter:2013xra,Berlin:2014cfa}.  In
contrast to most of the aforementioned probes, Higgs boson radiation
from an initial-state quark is Yukawa-suppressed. As a result, in a
potential signal the Higgs boson would be part of the interaction
producing the DM, providing unique insight into the structure of the
DM coupling to SM particles. Recently, the ATLAS Collaboration has
published such searches using $20.3\,\ifb$ of proton--proton collision
data at $\sqrt{s} = 8\,\tev$, exploiting the Higgs boson decays to two
photons or a pair of bottom quarks~\cite{Aad:2015yga,Aad:2015dva}.

This Letter presents an update on the search for $h+\MET$, where the
Higgs boson decays to a pair of bottom quarks ($h\rightarrow
b\bar{b}$), using $3.2\,\ifb$ of $pp$ collision data collected by the
ATLAS detector at a centre-of-mass energy of 13 TeV during 2015.
The results are interpreted in the context of simplified models of DM,
characterised by a minimal particle content and the corresponding
renormalisable interactions~\cite{Abdallah:2015ter}.

Many simplified models of DM production contain a massive particle
which can be a vector, an axial-vector, a scalar or a pseudoscalar,
and mediates the interaction between DM and Standard Model
particles. In this search, simplified models involving a vector
mediator are considered following the recommendation in
Ref.~\cite{Abercrombie:2015wmb}.

In the first model~\cite{Carpenter:2013xra}, a vector
mediator, \Zprime, is exchanged in the $s$-channel, radiates the Higgs
boson and decays into two DM particles. A diagram for this process is
shown in Figure~\ref{fig:vecMed}. The vector mediator has an
associated baryon number $B$ , which is assumed to be gauge invariant
under $U(1)_{B}$ thus allowing it to couple to
quarks~\cite{delAguila:1985hkb}. This symmetry is spontaneously broken
to generate the \Zprime\ mass. However, there is no \Zprime\ coupling
to leptons as such couplings are tightly constrained by dilepton
searches.  Finally, the dark-matter candidate carries a baryon number,
which allows it to couple to quarks through the \Zprime. The
parameters of this model are as follows: the coupling of \Zprime\ to
dark matter ($g_{\chi}$); the coupling of \Zprime\ to quarks
($g_{q}$); the coupling of \Zprime\ to the SM Higgs boson ($g_{Z'}$);
the mixing angle between the baryonic Higgs boson, introduced in the
model to generate the $Z'$~mass, and the SM Higgs boson
($\sin\theta$); the \Zprime~mass ($m_{Z'}$); and the DM particle mass
($m_\chi$).

In the second model, apart from the vector mediator, the SM is
extended by an additional Higgs field doublet, resulting in five
physical Higgs bosons~\cite{Berlin:2014cfa}: a light scalar $h$
associated with the observed Higgs boson, a heavy scalar $H$, a
pseudoscalar $A$, and two charged scalars $H^\pm$. The vector mediator
is produced resonantly and decays as $\Zprime\to hA$ in a Type-II
two-Higgs-doublet model (2HDM)~\cite{Branco:2011iw}. The pseudoscalar
$A$ subsequently decays into two DM particles with a large branching
ratio. A diagram for this process is shown in Figure~\ref{fig:2hdm}.
To define the model, the ratio of the up-- and down--type vacuum
expectation values, $\tan\beta$, must be specified along with the $Z'$
gauge coupling, $g_{Z}$, the DM particle mass, $m_{\chi}$, and
the \Zprime\ and $A$ masses, $m_{\Zprime}$ and $m_A$, respectively.
The results presented are for the alignment limit, in which the
$h$--$H$ mixing angle $\alpha$ is related to $\beta$ by $\alpha
= \beta - \pi/2$. Only regions of parameter space consistent with
precision electroweak constraints~\cite{Agashe:2014kda} and with
constraints from direct searches for dijet
resonances~\cite{Aaltonen:2008dn, CMS:2012yf, Khachatryan:2015sja} are
considered. As the $A$ boson is produced on-shell and decays into DM, the mass of
the DM particle does not affect the kinematic properties or
cross-section of the signal process if it is below half of the $A$
boson mass. Hence, the \Zprime-2HDM model is interpreted in the
parameter spaces of \Zprime\ mass ($m_{\Zprime}$), $A$ mass ($m_A$)
and $\tan{\beta}$.

\begin{figure}[h]
\centering
\subfigure[\label{fig:vecMed}]{\includegraphics[width=0.3\textwidth]{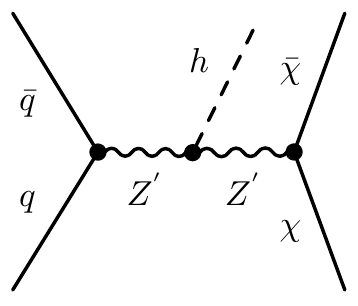}}
\hspace{2cm}
\subfigure[\label{fig:2hdm}]{\includegraphics[width=0.3\textwidth]{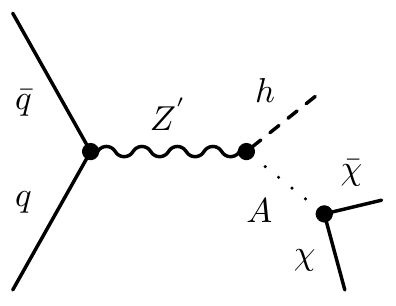}}
\caption{\label{fig:feynman} Diagrams showing the simplified models where \subref{fig:vecMed} a \Zprime\ decays to a 
pair of DM candidates $\chi\bar{\chi}$ after emitting a Higgs boson $h$, and where 
\subref{fig:2hdm} a \Zprime\ decays to a Higgs boson $h$ and the pseudoscalar $A$ of a two--Higgs--doublet model, and the latter decays to a pair of DM candidates $\chi\bar{\chi}$.}
\end{figure}

%% file: sec-detector.tex
ATLAS is a multi-purpose particle physics detector~\cite{Aad:2008zzm}
at the LHC, with an approximately forward-backward symmetric and
hermetic cylindrical geometry.\footnote{ATLAS uses a right-handed
coordinate system with its origin at the nominal interaction point
(IP) in the centre of the detector and the $z$-axis along the beam
pipe. The $x$-axis points towards the centre of the LHC ring, and the
$y$-axis points upwards. Cylindrical coordinates $(r, \phi)$ are used
in the transverse plane, $\phi$ is the azimuthal angle around the beam
pipe. The pseudorapidity $\eta$ is defined as $\eta = -\ln[\tan
(\theta/2)]$, where $\theta$ is the polar angle.  Finally, the angular
distance $\Delta R$ is defined as
$\sqrt{(\Delta\phi)^{2}+(\Delta\eta)^{2}}$.}  At its innermost part
lies the inner detector~(ID), immersed in a 2~T axial magnetic field
provided by a thin superconducting solenoid, consisting of silicon
pixel and microstrip detectors, which provide precision tracking in
the pseudorapidity range $|\eta| < 2.5$. It is complemented by a
transition radiation tracker providing tracking and particle
identification information for $|\eta| < 2.0$.  Between Run 1 and Run
2 of the LHC, the pixel detector was upgraded by the addition of a new
innermost layer~\cite{Capeans:1291633} that significantly improves the
identification of heavy-flavour
jets~\cite{ATL-PHYS-PUB-2015-022,ATL-PHYS-PUB-2015-039}.  The solenoid
is surrounded by sampling calorimeters: a lead/liquid-argon (LAr)
electromagnetic calorimeter for $|\eta| < 3.2$ and a
steel/scintillator tile hadronic calorimeter for $|\eta| <
1.7$. Additional LAr calorimeters with copper and tungsten absorbers
provide coverage up to $|\eta|=4.9$. In the outermost part, air-core
toroids provide the magnetic field for the muon spectrometer. The
latter consists of three layers of gaseous detectors: monitored drift
tubes and cathode strip chambers for muon identification and momentum
measurements for $|\eta|<2.7$, and resistive-plate and thin-gap
chambers for triggering up to $|\eta|=2.4$. A two-level trigger
system, custom hardware followed by a software-based level, is used to
reduce the event rate to about 1\,kHz for offline storage.

%% file: sec-data-mc.tex
The data sample used in this search, collected during normal operation
of the detector, corresponds to an integrated luminosity of
$3.2\,\ifb$. The primary data sample is selected using a
calorimeter-based \Etmiss\, trigger with a threshold of 70~GeV. The
trigger efficiency for signal events selected by the offline analysis
is about 90\% for events with \Etmiss\ of 150~GeV and reaches 100\%
for events with \Etmiss\ larger than 200~GeV.

Signal samples are generated at tree level with {\sc MadGraph}5\_aMC@NLO
2.2.3~\cite{Alwall:2014hca}, interfaced to {\sc Pythia}
8.186~\cite{pythia8} using the NNPDF2.3 parton distribution function
(PDF) set~\cite{Ball:2012cx} and the A14 parameter
tune~\cite{ATL-PHYS-PUB-2014-021} for parton showering, hadronisation,
underlying-event simulation, and for simulation of the Higgs boson
decay to a pair of bottom quarks. For the vector-mediator simplified
models, signals are generated with mediator mass between 10 and
$2000$~GeV and DM particle mass between 1 and $1000$~GeV. The event
kinematics are largely independent of the other parameters of the
model, and thus the same values of these parameters are chosen
following the recommendations in Ref.~\cite{Abercrombie:2015wmb}:
$g_\mathrm{\chi} = 1.0$, $g_\mathrm{q} = 1/3$, $g_\mathrm{Z'} =
m_\mathrm{Z'}$, $\sin\theta = 0.3$. For the \Zprime-2HDM model,
$pp\to\Zprime\to Ah\to \chi\bar{\chi}h$ samples are produced
with \Zprime\ mass values between 600 and $1000$~GeV, $A$ mass values
between 300 and $800$~GeV (where kinematically allowed), and a DM mass
value of $100$~GeV. The other parameters chosen for this model are
taken to be $\tan\beta$ = 1.0 and $g_{Z}$ = 0.8.

Higgs boson production in association with a $W$ or $Z$ vector boson,
$Vh$, is modelled using {\sc Pythia} 8.186 and the NNPDF2.3 PDF
set. The samples are normalised using the SM total cross-sections
calculated at next-to-leading order (NLO)~\cite{Han:1991ia} and
next-to-next-to-leading order (NNLO)~\cite{Brein:2003wg} in QCD for
$Wh$ and $Zh$, respectively, and include NLO electroweak
corrections~\cite{Ciccolini:2003jy}. In all cases, the Higgs boson
mass is set to 125~GeV.

Simulated samples of vector boson production in association with jets,
$W/Z$+jets, where the $W$ or $Z$ bosons decay in all leptonic decay
modes, are generated using \SHERPA 2.1.1~\cite{Gleisberg:2008ta},
including $b$- and $c$-quark mass effects, and the CT10 PDF
set~\cite{Lai:2010vv}. Matrix elements are calculated for up to two
partons at NLO and four partons at LO using the
Comix~\cite{Gleisberg:2008fv} and OpenLoops~\cite{Cascioli:2011va}
matrix element generators and merged with the \SHERPA parton
shower~\cite{Schumann:2007mg} using the ME+PS@NLO
prescription~\cite{Hoeche:2012yf}. The cross-sections are determined
at NNLO~\cite{Melnikov:2006kv} in QCD.  Furthermore, these backgrounds
are split into different components according to the true flavour of
the two jets that are used to identify the flavor of the reconstructed
Higgs boson candidate, as described in
Section~\ref{sec:eventselection}: $l$ denotes a light quark ($u,d,s$)
or a gluon and the heavy quarks are denoted by $c$ and $b$. This
division is performed to allow accurate modelling of the $W/Z$+
heavy-flavour backgrounds in the combined fit described in
Section~\ref{sec:Results}.

Diboson production modes, including $ZZ$, $WW$, and $WZ$ processes,
with one boson decaying hadronically and the other leptonically are
simulated using the \SHERPA 2.1.1 generator with the CT10 PDF
set. They are calculated for up to one ($ZZ$) or zero ($WW/WZ$)
additional partons at NLO and up to three additional partons at LO
using the Comix and OpenLoops matrix element generators and merged
with the \SHERPA parton shower using the ME+PS@NLO prescription. Their
cross-sections are determined by the generator at NLO.

The \ttbar and single-top-quark backgrounds are generated with {\sc
PowhegBox}~\cite{Powheg} using the CT10 PDF set. It is interfaced with
{\sc Pythia} 6.428~\cite{Sjostrand:2006za} to simulate parton
showering, fragmentation, and the underlying event, for which the
CTEQ6L1 PDF set~\cite{cteq6l1} and the Perugia 2012 parameter
tune~\cite{Skands:2010ak} are used. The \ttbar cross-section is
determined at NNLO in QCD and next-to-next-to-leading logarithms
(NNLL) for soft gluon radiation~\cite{Czakon:2013goa}, while the
single-top-quark cross-sections are fixed to those in
Refs.~\cite{Kidonakis:2011wy, Kidonakis:2010tc, Kidonakis:2010ux}. A
top-quark mass of 172.5~GeV is used throughout.

The simulated event samples are processed with the detailed ATLAS
detector simulation~\cite{Aad:2010ah} based on {\sc
Geant4}~\cite{Geant4}.  Effects of multiple proton--proton
interactions (pile-up) as a function of the instantaneous luminosity
are taken into account by overlaying simulated minimum-bias events
generated with {\sc Pythia}8.186 with the A2 tune~\cite{ATLAS:2012uec}
and MSTW2008LO PDF set~\cite{PDF-MSTW2008} onto the hard-scattering
process, such that the distribution of the average number of
interactions per bunch crossing in the simulated event samples matches
that in the data.

%% file: sec-object.tex
Proton--proton collision vertices are reconstructed using ID tracks
with $\pt> 0.4\,\gev$. The primary vertex is defined as the vertex
with the highest $\Sigma (p_\mathrm{T}^{\mathrm{track}} )^2$. Each
event is required to have at least one vertex reconstructed from at
least two tracks.

Muon candidates are identified by matching tracks found in the ID to
either full tracks or track segments reconstructed in the muon
spectrometer, and are required to satisfy the {\it loose} muon
identification quality criteria~\cite{Aad:2016jkr}. Electron
candidates are identified as ID tracks that are matched to a cluster
of energy in the electromagnetic calorimeter. Electron candidates must
satisfy a likelihood-based identification
requirement~\cite{ATLAS:2014wga} based on shower shape and track
selection criteria, and are selected using the {\it loose} working
point.  Both the muons and electrons are required to originate from
the primary vertex, to have $\pt>7\,\gev$, and to lie within
$|\eta|<2.5$ for muons and $|\eta|<2.47$ for electrons. They are
further required to be isolated using requirements on the sum of $\pt$
of the tracks within a cone around the lepton direction. The cone size
and the requirements are varied as a function of the lepton $\pt$ to
obtain an efficiency that is fixed as a function of $\pt$ such that a
99\% efficiency for prompt leptons is retained across a broad
kinematic range.

Jets are reconstructed in two categories, small-radius (small-$R$) and
large-radius (large-$R$) jets. In both cases, the jets are
reconstructed from topological clusters of calorimeter cells using
the \akt\ jet clustering algorithm~\cite{Cacciari:2008gp}.  In the
case of small-$R$ jets, a radius parameter of $R = 0.4$ is used and
the effects of pile-up are corrected for by a technique based on jet
area~\cite{Cacciari:2008gn}. In the case of large-$R$ jets, a radius
parameter of $R = 1.0$ is used and the jet trimming
algorithm~\cite{Ellis1993,Krohn2010} is applied to minimise the impact
of energy depositions due to pile-up and the underlying event. This
algorithm reconstructs subjets within the large-$R$ jet using the
$k_{\mathrm t}$ algorithm~\cite{Catani:1993hr} with radius parameter
$R_{\rm sub} = 0.2$ and removes any subjet with $\pt$ less than 5\% of
the large-$R$ jet \pt. The jet energy scale, and also in the case of
large-$R$ jets the jet mass scale, is calibrated using \pt- and
$\eta$-dependent factors determined from simulation, with small-$R$
jets receiving further calibrations using {\it in situ}
measurements~\cite{Aad:2011he}. Small-$R$ jets within the ID
acceptance, $|\eta|<2.5$, are called {\it central} in the following
and are required to satisfy $\pT>20\,\gev$. Those with
$2.5<|\eta|<4.5$ are called {\it forward} and are required to satisfy
$\pT>30\,\gev$. To reduce the effects of pile-up in small-$R$ jets
with $\pt<50\,\gev$ and $|\eta|<2.5$, a significant fraction of the
tracks associated with each jet must have an origin compatible with
the primary vertex, as defined by the jet vertex
tagger~\cite{ATLAS:2014cva}. Furthermore, small-$R$ jets are removed
if they are within a $\Delta R$ = 0.2 cone around an electron
candidate.  Large-$R$ jets are required to satisfy $\pT>250\,\gev$ and
$|\eta|<2.0$.

Track jets are built from tracks using the \akt\ algorithm with
$R=0.2$. Track jets with $\pT>10\,\gev$ and $|\eta|<2.5$ are selected
and are matched by ghost-association~\cite{Cacciari:2007fd} to
large-$R$ jets. Small-$R$ jets and track jets containing $b$-hadrons
are identified--``$b$-tagged''--using a boosted decision tree that
combines information about the impact parameter and reconstructed
secondary vertices of the tracks associated with these
jets~\cite{ATL-PHYS-PUB-2015-022,ATL-PHYS-PUB-2015-039,Aad:2015ydr}.
A working point is used which achieves an average efficiency of 70\%
in identifying small-$R$ calorimeter jet (track jet) containing a
$b$-hadron with misidentification probabilities of $\sim 12$ (18)\%
for charm-quark jets and $\sim 0.2$ (0.6)\% for light-flavour jets, as
determined in a simulated sample of
\ttbar~events. Track jets have higher misidentification probabilities due to the smaller radius parameter used.

The missing transverse momentum, $\vecMET$, is defined as the negative
vector sum of the transverse momenta of the calibrated physics objects
(electrons, muons, small-$R$ jets), with unassociated energy
depositions, referred to as the soft-term, accounted for using ID
tracks with $p_{\mathrm{T}}$ > 0.5
GeV~\cite{ATL-PHYS-PUB-2015-023,Aad:2012re}. Furthermore, a
track-based missing transverse momentum vector, $\vecMPT$, is
calculated as the negative vector sum of the transverse momenta of
tracks with $|\eta| < 2.5$, consistent with originating from the
primary vertex.\footnote{Throughout this search, the magnitude of
$\vec{E}_{\mathrm{T}}^{\mathrm{miss}}$ is referred to as
$E_{\mathrm{T}}^{\mathrm{miss}}$ and the magnitude of $\vecMPT$ is
referred to as $\MPT$. Only when the directionality is necessary does
the notation use the vector symbol.}

%% file: sec-analysis.tex
\section{Event selection}
\label{sec:eventselection}

For an event to be considered in the search, it is required to have
$\Etmiss>150\,\gev$, $\MPT>30\,\gev$, and no identified, isolated
muons or electrons. This is referred to as the {\it zero-lepton
  region}.

Events with $\Etmiss$ less than $500\,\gev$ are considered in the {\it
  resolved region}. First, this set of events is required to have at
least two central small-$R$ jets.  Following this selection, the
reconstructed small-$R$ jets are ranked as follows.  First, the
central jets are divided into two categories, those that are
$b$-tagged and those that are not.  Each of these samples of jets are
ordered in decreasing $\pt$.  The ordered set of $b$-tagged jets is
considered with the highest priority, while those that are central but
not $b$-tagged are considered with second priority, and finally any
forward jets, ordered in decreasing $\pt$, are considered last.  The
two most highly ranked jets are used to reconstruct the Higgs boson
candidate, \hreco, and therefore cannot contain forward
jets. Furthermore, at least one of the jets constituting
$h_{\mathrm{r}}$ must satisfy $\pt>45\,\gev$.  Finally, events are
divided into three categories based on the number of central jets that
are $b$-tagged being either zero, one, or two $b$-tagged central jets.
To achieve a high \Etmiss\ trigger efficiency, events are retained if
the scalar sum of the \pt~of the three leading jets is greater than
150 GeV. This requirement is lowered to 120 GeV if only two central
small-$R$ jets are present.

Additional selections are applied to further suppress the multijet
background. Specifically, to reject events with \MET~due to
mismeasured jets a requirement is placed on the minimum azimuthal
angle between the direction of the \Etmiss~and each of the jets,
$\min\left(\Delta\phi\left(\vecMET,{\rm jets}\right)\right)>20^\circ$,
for the three highest-ranked jets. Furthermore, the azimuthal angle
between the \vecMET~and the \vecMPT,
$\Delta\phi\left(\vecMET,\vecMPT\right)$, is required to be less than
$90^\circ$, to suppress events with misreconstructed missing
transverse momentum. The Higgs boson candidate is required to be well
separated in azimuth from the missing transverse momentum by requiring
$\Delta\phi\left(\vecMET,\hreco\right)>120^\circ$. Finally, to reject
back-to-back dijet production, the azimuthal opening angle of the two
jets forming the Higgs boson candidate is required to be
$\Delta\phi\left(j^{1}_{\hreco},j^{2}_{\hreco}\right)< 140^\circ$.

The DM signal is expected to have large \MET, whereas the background
is expected to be most prominent at low \MET. Therefore, to retain
signal efficiency while preserving the increased sensitivity of the
high \MET region, events in the resolved region are separated into
three categories based on the reconstructed \MET: 150--200~\gev,
200--350~\gev, and 350--500~\gev.

In the {\it merged region}--composed of events with $\Etmiss$ in
excess of $500\,\gev$--the presence of at least one large-$R$ jet is
required, associated with at least two track
jets~\cite{ATL-PHYS-PUB-2015-035}, and the highest \pt large-$R$ jet
is taken as the reconstructed Higgs candidate. In an analogous way to
the resolved region, the events are classified based on the number of
$b$-tagged track jets associated with the large-$R$ jet into three
categories with zero, one, and two or more $b$-tags.

The combined selection of both the resolved and merged selections in
the signal region with two or more $b$-tags yields a signal acceptance
times efficiency ranging between 5 and 30\%. The primary change in
the signal acceptance is due to the choice of masses (e.g. $m_{Z'}$
and $m_{A}$) in the point of parameter space being probed.

The search is performed by implementing a shape fit of the
reconstructed dijet mass ($m_{\mathrm{jj}}$) or single large-$R$ jet
mass ($m_\mathrm{J}$) distribution.  After event selection, the energy
calibration of the $b$-tagged jets is improved as follows.  The
invariant mass of the candidate is corrected~\cite{Aad:2014xzb} if a
muon is identified within $\Delta R$ = 0.4 of a $b$-tagged small-$R$
jet, or within $\Delta R$ = 1.0 of the large-$R$ jet.  The
four-momentum of the closest muon in $\Delta R$ within a jet is added
to the calorimeter-based jet energy after removing the energy
deposited in the calorimeter by the muon (muon-in-jet correction).
Additionally, a simulation-based jet-\pt-dependent
correction~\cite{Aad:2014xzb} is applied in the case of $b$-tagged
small-$R$ jets to improve the signal resolution of the reconstructed
Higgs mass peak. Events consistent with a DM signal would have a
reconstructed mass near the Higgs boson mass, thereby allowing the
sidebands to act as a natural control region to further constrain the
backgrounds estimated from dedicated $W/Z+\mathrm{jets}$ and \ttbar
control regions and the multijet estimates described in
Section~\ref{sec:background}.

\section{Background estimation}
\label{sec:background}
The background is mainly composed of SM $W/Z+\mathrm{jets}$ and \ttbar
events, which constitute 15--65\% and 45--80\% of the total
background, respectively, depending on the \MET~value. The model for
these backgrounds is constrained using two dedicated control regions.
Other backgrounds, including diboson, $Vh$, and single top-quark
production, constitute less than 15\% of the total background and the
estimation is modelled using simulated event samples. The contribution
from multijet events arises mainly from events containing jets
containing semi-muonic decays of $b$-hadrons. It constitutes less than
2\% of the background in the resolved region and is negligibly small
in the merged region, and is estimated using a data-driven technique.

In addition to the zero-lepton region, which serves as a control
region to constrain the $Z+\mathrm{jets}$ background in the
zero-$b$-tag case and via the reconstructed mass sidebands that enter
in the fit as described in Section~\ref{sec:Results}, two dedicated
control regions are used to constrain the main $W/Z+\mathrm{jets}$ and
\ttbar~backgrounds. These control regions are defined based on the
number of leptons and $b$-tags in the event and are orthogonal to each
other and to the signal region.

The {\it one-muon control region} is designed to constrain the
$W+\mathrm{jets}$ and \ttbar~backgrounds. Events are selected using
the \MET trigger and are required to have exactly one muon candidate
and no electron candidates. Furthermore, the full signal region
selection is applied after modifying the \Etmiss~observable to mimic
the behaviour of such events that contaminate the signal region by
adding the \pt of the reconstructed muon to the \MET. As in the signal
region, these events are divided into exclusive regions based on the
number of $b$-tags. This division naturally separates \ttbar from
$W+\mathrm{jet}$ events.

The {\it two-lepton control region} is used to constrain the
$Z+\mathrm{jets}$ background contribution. Events are collected using
a single-electron or single-muon trigger and selected by requiring
exactly one electron pair or muon pair. Of these two leptons, one is
required to have $\pt>25\,\gev$. The electron (muon) pair must have
an invariant mass $83 < m_{\ell\ell} < 99\,\gev$ ($71 < m_{\ell\ell} <
106\,\gev$). In the muon channel, where a larger mass window is used,
an opposite-charge requirement is also applied. Furthermore, the
missing transverse momentum significance, defined as the ratio of
\Etmiss~to the square root of the scalar sum of lepton and jet \pt~in
the event, is required to be less than $3.5\,\gev^{1/2}$ in order to
reject \ttbar background. In this control region, the transverse
momentum of the dilepton system, $\pt^V$, is used--instead of
\Etmiss--to match the division of the resolved and merged regions and
the categorisation of the resolved events. Other than the above, the
event selection and Higgs boson candidate requirements are the same as
in the signal region.

The multijet background for the resolved analysis is determined using
a data-driven method. A sample of events selected to satisfy the
analysis trigger, \MPT~requirement, and inverted ${\rm
  min}(\Delta\phi(\vecMET ,{\rm jets}))$ requirement, is used to
provide multijet templates of all the distributions relevant to the
analysis. These templates are normalised by a fit to the distribution
of the number of small-$R$ jets that contain a muon in the nominal
selection.  The fit is performed separately for each $b$-tag category.
Since agreement is found between the categories the average
normalisation scale factor is used. In the merged region, it was
found that the requirement of high \MET suppresses the multijet
background to a negligible level. Therefore it is not included as a
background in the search.

%% file: sec-systematics.tex
\section{Systematic uncertainties}
\label{sec:ResolvedSyst}

The most important experimental systematic uncertainties arise from
the determination of the $b$-tagging efficiency and mistag rate, the
luminosity determination and uncertainties associated with the
calibration of the scale and resolution of the jet energy and mass.
The uncertainties in the small-$R$ jet energy scale have contributions
from {\it in situ} calibration studies, from the dependence on pile-up
activity and on flavour composition of jets, and from the changes of
the detector and run conditions between Run 1 and
Run~2~\cite{Aad:2014bia,ATL-PHYS-PUB-2015-015}.  The uncertainty in
the scale and resolution of large-$R$ jet energy and mass are
evaluated by comparing the ratio of calorimeter-based to track-based
measurements in dijet data and
simulation~\cite{ATL-PHYS-PUB-2015-033}. The $b$-tagging efficiency
uncertainty arises mainly from the uncertainty in the measurement of
the efficiency in $t\bar{t}$ events~\cite{Aad:2015ydr,ATLAS:2016wlr}.

Other experimental systematic uncertainties with a smaller impact are
those in the lepton energy and momentum scales, and lepton
identification and trigger
efficiencies~\cite{Aad:2016jkr,Aad:2014fxa,Aad:2014nim}.  An
uncertainty in the \Etmiss\ soft-term resolution and scale is taken
into account~\cite{ATL-PHYS-PUB-2015-023}, and uncertainties due to
the lepton energy scales and resolutions, as well as reconstruction
and identification efficiencies, are also considered, although they
are negligible.  The uncertainty in the integrated luminosity amounts
to 2.1\%, and is derived following a methodology similar to that
detailed in Ref.~\cite{Aad:2013ucp}.

Uncertainties are also taken into account for possible differences
between data and the simulation modelling used for each process. The
\SHERPA $W+\mathrm{jets}$ and $Z+\mathrm{jets}$ background modelling
is studied in the one and two lepton control regions, respectively, as
a function of $\pt$ of the vector boson, the mass $m_{\mathrm{jj}}$ or
$m_{\mathrm{J}}$ and the azimuthal angle difference $\Delta
\phi_{\mathrm{jj}}$ between the small-$R$ jets used to reconstruct the
Higgs in the resolved region. The shape of the data distributions is
described by the simulation with no indication that a correction is
needed. A shape uncertainty in these variables is derived,
encompassing the data/simulation differences. An uncertainty in the
\SHERPA description of the flavour composition of the jets in these
backgrounds is derived by comparing to \Madgraph. The top-quark
background modelling is studied in the dedicated one lepton control
region, and in a two lepton control region using $e\mu$ pairs. Both
the $\pt$ and mass of the two small-$R$ jet system are studied.  A
systematic uncertainty is derived based on the data/simulation
comparison in these regions.

The normalisations of the $W+b\bar{b}$, $Z+b\bar{b}$, and $t\bar{t}$
contributions are determined directly from the data by leaving them as
free parameters in the combined fit. The normalisations of the other
$W/Z+\mathrm{jets}$ background contributions are obtained from theory
predictions, with assigned normalisation uncertainties of 10\% for
$W/Z+l$, 30\% for $W/Z+cl$ and a 30\% uncertainty is applied to the
relative normalisation between $W/Z+bc/bl/cc$ to $W/Z+b\bar{b}$.  In
addition, the following normalisation uncertainties are assigned to
the background processes: 4\% for single-top in the $s$- and $t$-
channels, 7\% for single-top in the $Wt$-
channel~\cite{Aliev:2010zk,Kant:2014oha}, and 50\% for associated
$(W/Z)h$~\cite{Aad:2014xzb,Chatrchyan:2013zna} production.  The
sources of uncertainty considered for the cross-sections for the
diboson production ($WW$, $WZ$ and $ZZ$) are the renormalisation and
factorisation scales, the choice of PDFs and parton-shower and
hadronisation model. The multijet contribution is estimated from data
and is assigned a 50\% uncertainty. Uncertainties arising from the
size of the simulated event sample are also taken into account.

Uncertainties in the signal acceptance from the choice of PDFs, from
the choice of factorisation and renormalisation scales, and from the
choice of parton-shower and underlying-event tune have been taken into
account in the analysis. These are typically $<10\%$ each, although
they can be larger for regions with low acceptance at either low or
high \Etmiss\ depending on the model and the choice of masses.  In
addition, uncertainties arising from the limited number of simulated
events have been taken into account.

The contribution of the various sources of uncertainty for an example
production scenario is given in Table~\ref{tab:systs}.

\begin{table*}[htb]
\centering
\begin{tabular}{|l|c|}
\hline
Source of Uncertainty & Impact [\%] \\
\hline
Total & 23.0\\
Statistical & 20.5\\
Systematic & 10.3\\
\hline
\multicolumn{2}{c}{Experimental Uncertainties}\\
\hline
$b$-tagging & 6.6\\
Luminosity &  4.4\\
Jets+$E_{\mathrm{T}}^{\mathrm{miss}}$ & 2.8\\
Leptons & 0.4\\
\hline
\multicolumn{2}{c}{Theoretical and Modelling Uncertainties}\\
\hline
Top & 5.1\\
$Z$+jets & 3.4\\
Signal & 2.6\\
$W$+jets & 1.5\\
Diboson & 0.6\\
Multijet & 0.5\\
$Vh$ $(h \rightarrow b\overline{b})$ & 0.4\\
\hline
\end{tabular}
\caption{\label{tab:systs} The percentage impact of the various
  sources of uncertainty on the expected production cross-section 
  for the signal in the vector-mediator model
  with $m_{Z'}=2000$ GeV and $m_\chi=1$ GeV, normalised to a cross section of 0.1 pb.}
\end{table*}

%% file: sec-results.tex
\label{sec:Results}

\begin{figure}[h]
\centering
\includegraphics[width=0.45\textwidth]{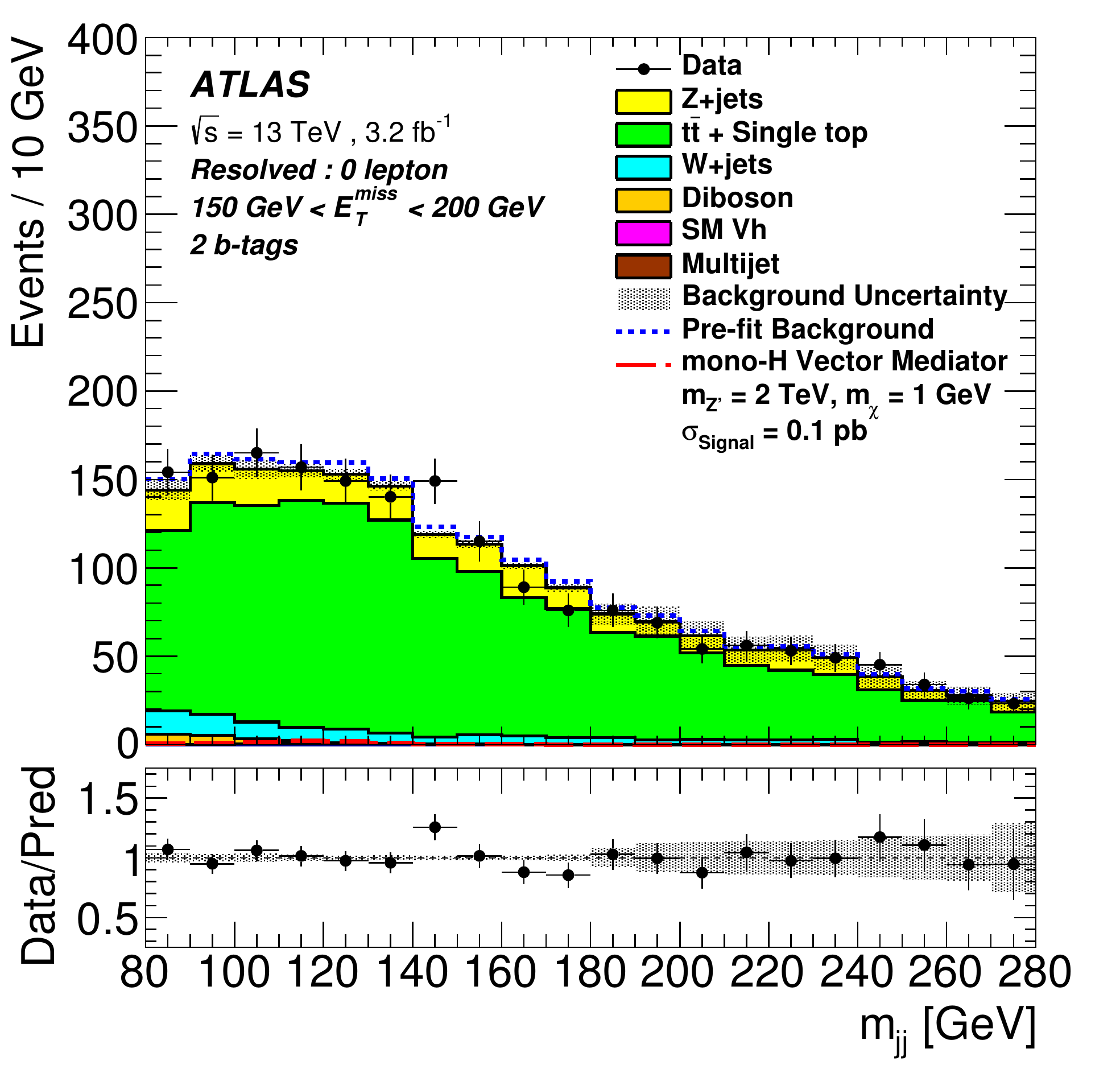}
\includegraphics[width=0.45\textwidth]{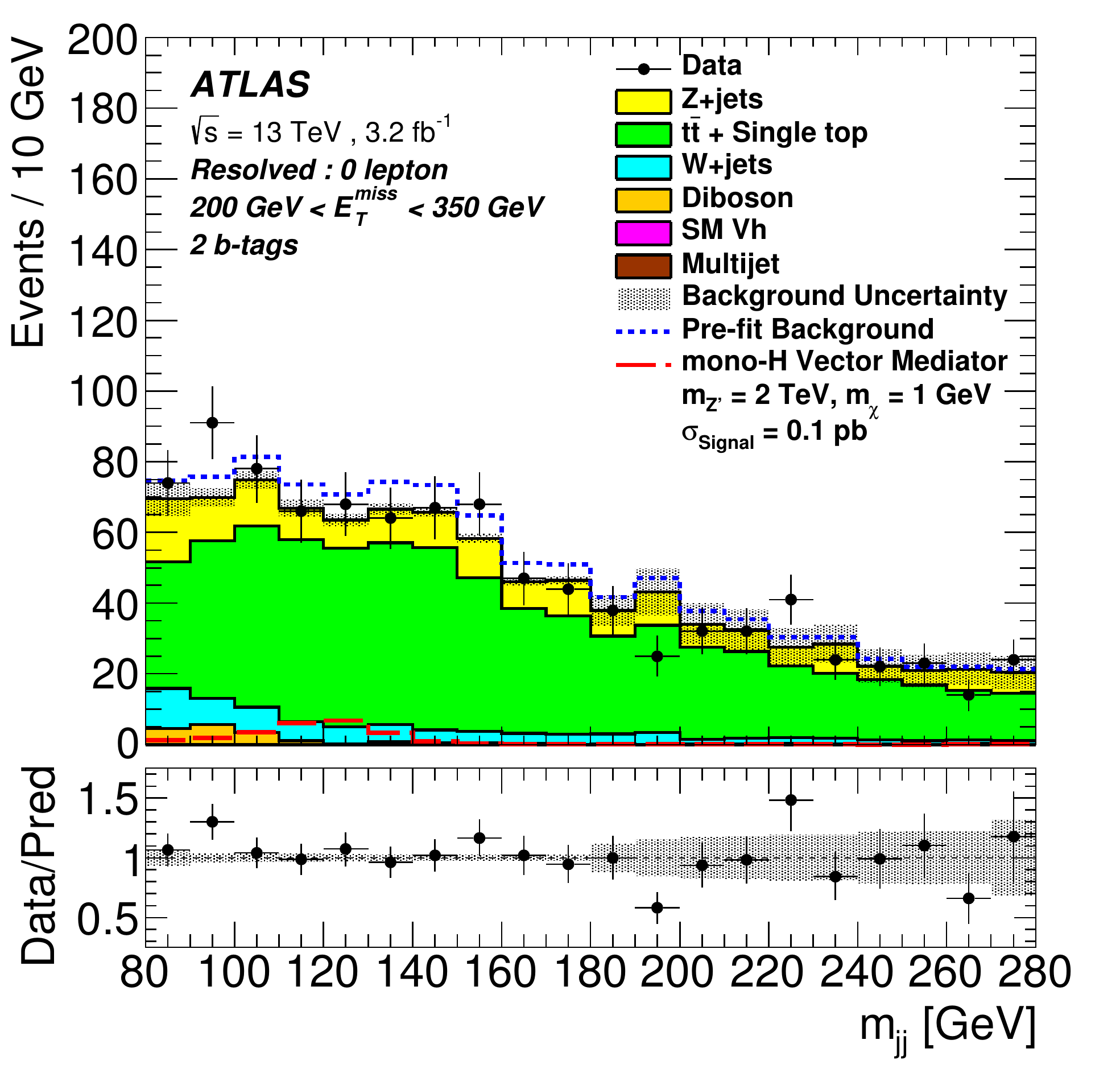}
\includegraphics[width=0.45\textwidth]{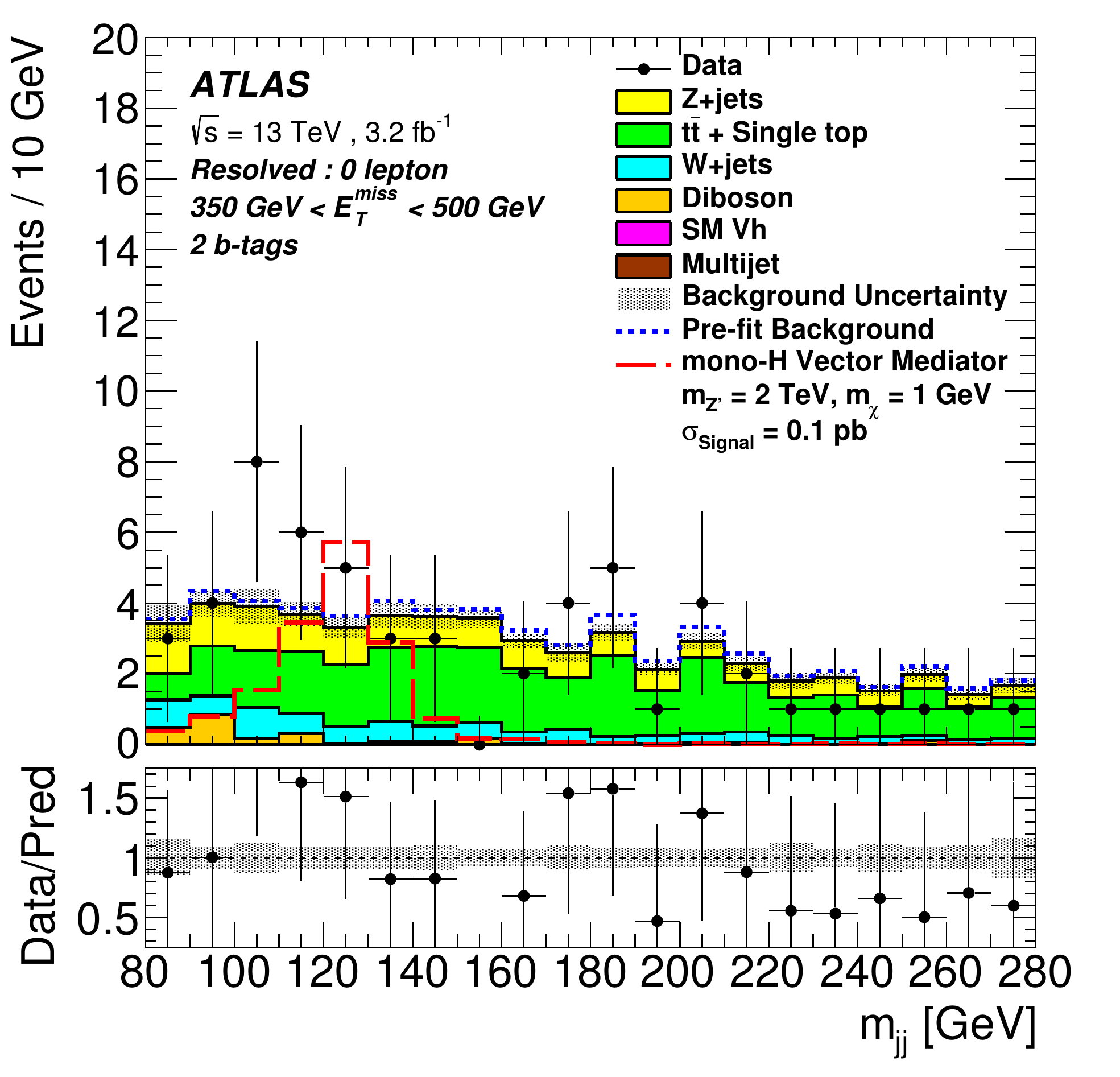}
\includegraphics[width=0.45\textwidth]{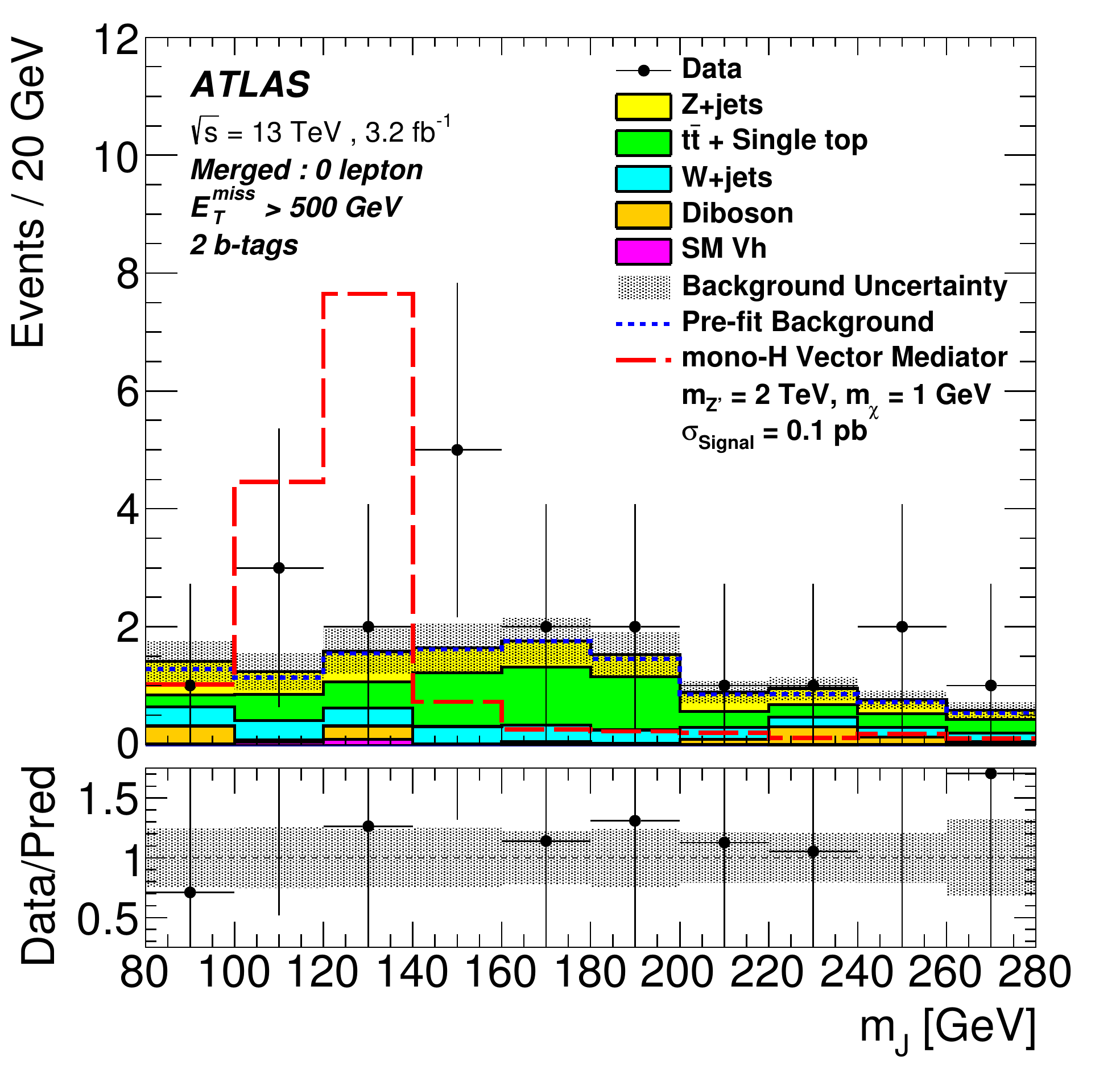}
\caption{The reconstructed dijet and single jet invariant mass distribution in the resolved and the merged signal regions for the case
 where two $b$-tags have been identified for the four kinematic
 regions. The Standard Model background expectation is shown before
 (after) the profile likelihood fit by the dashed blue line (solid
 histograms) with the bottom panel showing the ratio of the data to
 the predicted background after the combined fit with no signal
 included.  For visual clarity the various components of the
 $W/Z+\mathrm{jets}$ ($b\bar{b}$, $bc$, $bl$, $c\bar{c}$, $cl$, $ll$)
 backgrounds have been merged and labelled $W+\mathrm{jets}$ and
 $Z+\mathrm{jets}$. The expected signal in the vector-mediator model
 with $m_{Z'}=2$~TeV and $m_\chi=1$~GeV, normalised with a
 cross-section of 0.1 pb, is also shown.
\label{fig:postfit_mBB}}
\end{figure}

\begin{figure}[h]
\centering
\includegraphics[width=0.49\textwidth]{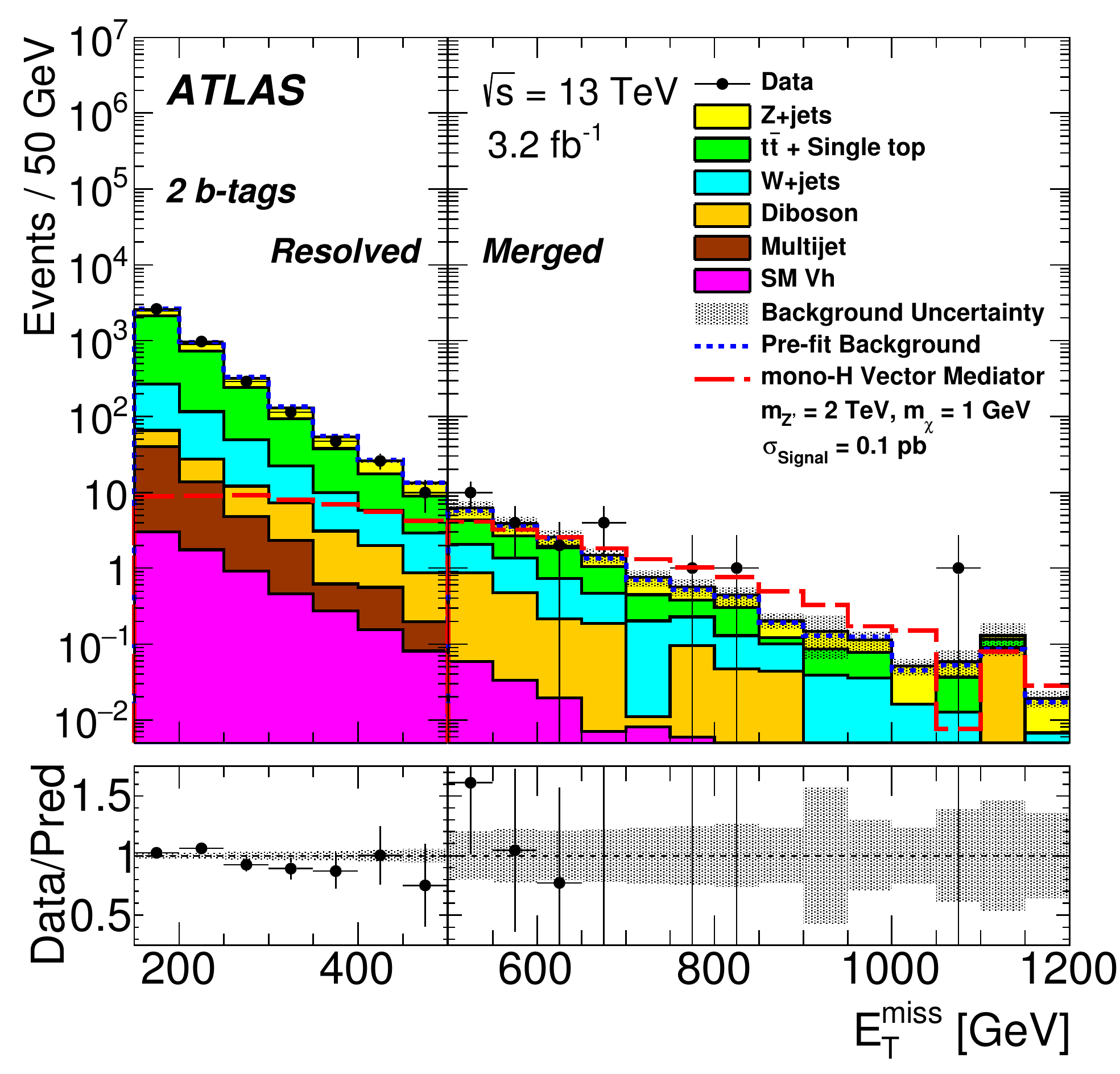}
\caption{The reconstructed \Etmiss\ distribution in the combined resolved and merged two-$b$-tag signal regions.
The Standard Model prediction is shown before (after) the profile
 likelihood fit by the dashed blue line (solid histograms) with the
 bottom panel showing the ratio of the data to the predicted
 background after the combined fit with no signal included.  For
 visual clarity the various components of the $W/Z+\mathrm{jets}$
 ($bb$, $bc$, $bl$, $cc$, $cl$, $ll$) backgrounds have been merged and
 labelled $W+\mathrm{jets}$ and $Z+\mathrm{jets}$. The multijet
 background is found to be negligible in the merged region.  The
 expected signal in the vector-mediator model with $m_{Z'}=2$~TeV and
 $m_\chi=1$~GeV, normalised with a cross-section of 0.1 pb, is also
 shown. \label{fig:postfit_MET}}
\end{figure}

Results are extracted by means of a profile likelihood fit to the
reconstructed invariant mass distribution of the dijet system or single-large-$R$-jet 
simultaneously in all signal and control regions. The normalisations of the major backgrounds are contrained
by the data in both the signal and control regions. The shapes of the background distributions are taken from
Monte Carlo simulations but can be modified  within the systematic errors listed in Section~\ref{sec:ResolvedSyst}.
The spectra entering the fit are those from the three selections associated with the number of leptons
with each of these regions divided into three categories based on the number of $b$-tags and four kinematic regions.
In the zero-lepton region, this division is based on \MET~while in the one- and two-lepton regions, it is based on
$p_{\mathrm{T}}(\mu,\MET)$ and $p_{\mathrm{T}}(\ell,\ell)$, respectively.  The shape information is not used in
the zero-$b$-tag distributions in order to simplify the fit.  This division is designed to isolate, and more effectively 
constrain, different backgrounds.  In particular, the $Z+\mathrm{jets}$ background normalisation is
constrained both by the sample of events containing two leptons and those containing zero leptons and zero $b$-tags. 
In addition, the set of events containing one lepton and zero $b$-tags constrains the $W+\mathrm{jets}$ normalisation
while those containing one or two $b$-tags constrain both the $W+\mathrm{jets}$ and $t\bar{t}$ normalisations. 
 The parameter of
interest in the fit is the signal yield, while all parameters
describing the systematic uncertainties and their correlations are
included in the likelihood function as nuisance parameters, with
Gaussian constraints, implemented using the framework described in
Refs.~\cite{Verkerke:2003ir,Moneta:2010pm}. The nuisance parameters
with the largest effect on the determination of the parameter of
interest are the flavour-tagging and jet systematic uncertainties,
together with the normalisation of the $t\bar{t}$ and $W+b\bar{b}$
backgrounds. The reconstructed Higgs boson candidate mass distribution
is shown in Figure~\ref{fig:postfit_mBB} in each of the \MET
categories for the set of events with two $b$-tags with the integrated
event yields shown in Table~\ref{tab:resolvedRslts}. Furthermore,
shown in Figure~\ref{fig:postfit_MET} is the \MET~distribution in the
signal region, noting that in the two portions of the spectrum, below
and above $\MET$ = 500 GeV, the requirements on the hadronic activity
are taken from the small-$R$ and large-$R$ jets, respectively. No
significant excess of events is observed above the background, with
the global significance of the deviation of the data from the
background-only prediction being 0.056.

\begin{table*}[!ht]
{
\footnotesize
\begin{center}
\begin{tabular}{|l|c|c|c||c|}
\hline
\MET & \multicolumn{3}{c||}{Resolved}
& Merged \\ \cline{2-5}
[GeV] & 150--200 & 200--350 & 350--500 & $>$500\\ \hline
\hline
$Z$ + jets                  & \phantom{0}259 $\pm$ 27   & 171  $\pm$ 13             & 14.6 $\pm$ 1.2              & \phantom{0}3.80 $\pm$ 0.44 \\
$W$ + jets                  & \phantom{00}95 $\pm$ 28   & \phantom{0}70   $\pm$ 22  & \phantom{0}7.5  $\pm$ 2.4   & \phantom{0}2.48 $\pm$ 0.71 \\
$t\bar{t}$ \& Single top      & 1444 $\pm$ 44             & 656  $\pm$ 25             & 30.8 $\pm$ 1.4              & \phantom{0}4.9  $\pm$ 0.9 \\
Multijet                      & \phantom{00}21 $\pm$ 10   & 11.0 $\pm$ 5.0            & \phantom{0}0.58 $\pm$ 0.27  & -- \\
Diboson	                      & \phantom{0}17.8 $\pm$ 1.6 & 18.7 $\pm$ 1.0            & \phantom{0}2.53 $\pm$ 0.22  & \phantom{0}1.20 $\pm$ 0.12 \\
${\text{SM}}$ $Vh$            & \phantom{00}2.8 $\pm$ 1.3 & \phantom{0}2.8  $\pm$ 1.4 & \phantom{0}0.46 $\pm$ 0.23  & \phantom{0}0.15 $\pm$ 0.08 \\
\hline                     
Total Bkg.                    & 1840 $\pm$ 33             & 930  $\pm$ 20             & 56.5 $\pm$ 2.1   & 12.5 $\pm$ 1.3 \\
\hline                      
Data                          & 1830                      & 942                       & 56               & 20 \\ 
\hline                     
Exp. Signal                   & \phantom{00}8.0 $\pm$ 0.8     & 24.5  $\pm$ 1.8             & 16.1 $\pm$ 1.2     & 14.9 $\pm$ 3.4  \\
\hline
\end{tabular}
\caption{The numbers of predicted background events following the profile likelihood fit for each background
process, the sum of all background components, and observed data yields in
the two $b$-tag  signal region  of the resolved and merged channels for each
\MET region. Statistical and systematic
uncertainties are combined. The uncertainties in the total background
take into account the correlation of systematic uncertainties among
different background processes. The expected signal in the
vector-mediator model with $m_{Z'}=2000$ GeV and $m_\chi=1$ GeV.
\label{tab:resolvedRslts}}
\end{center}
}
\end{table*}

\begin{figure}[h]
\centering
\subfigure[\label{fig:Zprimeplot}]{\includegraphics[width=0.45\textwidth]{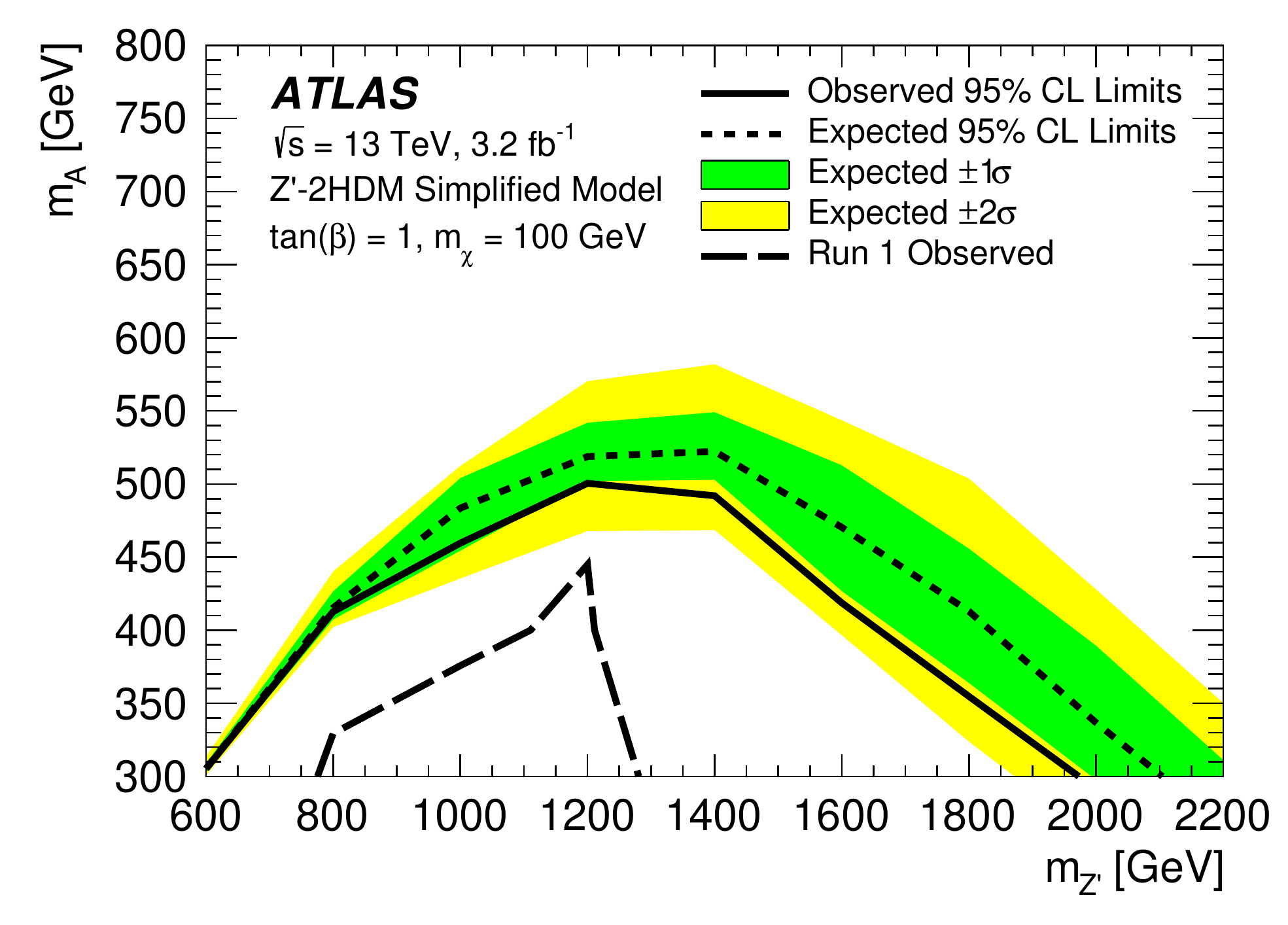}}
\subfigure[\label{fig:vectorplot}]{\includegraphics[width=0.45\textwidth]{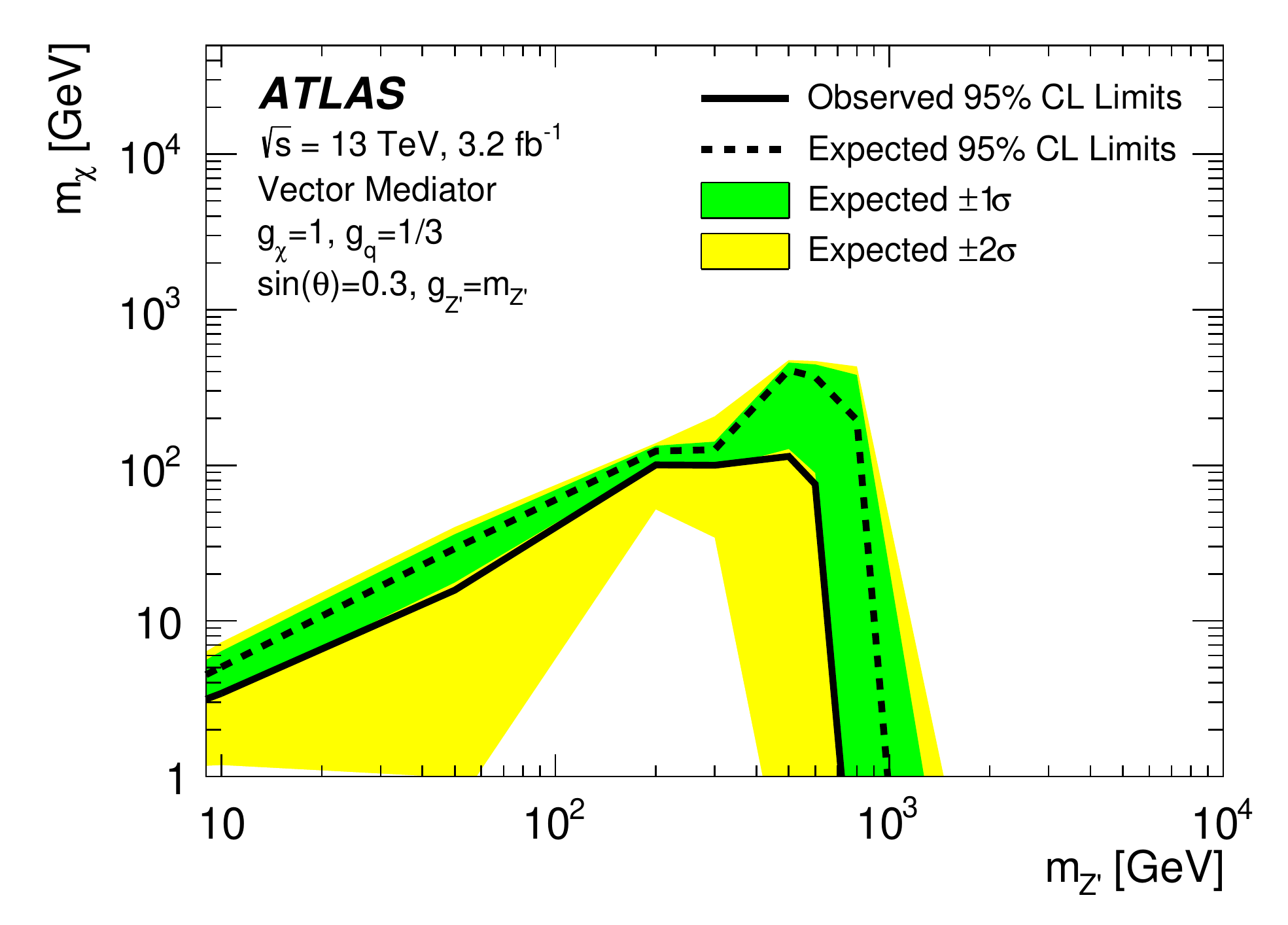}}
\caption{Exclusion contours for 
\subref{fig:Zprimeplot} the \Zprime-2HDM in the $(m_{\Zprime},m_{A})$ plane
for $\tan{\beta} = 1$ and $m_{\chi}=100\,\gev$ and
\subref{fig:vectorplot} the vector-mediator model in the
$(m_{Z'},m_\chi)$ plane for $\sin\theta=0.3$, $g_{\chi}=1$, $g_q=1/3$
and $g_{Z'}=m_{\Zprime}$. The expected limits are given by the dashed
lines, while the green and yellow bands indicate the $\pm 1 \sigma$
and $\pm 2 \sigma$ uncertainty bands, respectively. The observed
limits are given by the solid lines. The parameter space below the
limit contours are excluded at $95\%$ confidence level. Shown for
the \Zprime-2HDM exclusion is the observed limit from the Run 1 search
while no such exclusion is shown from Run 1 for the vector-mediator
model as it was not used for interpretation in the Run 1 ATLAS search.
\label{fig:simple_param_limits}}
\end{figure}

Upper limits on the production cross-section for the process times
branching ratio of the Higgs boson decaying to two bottom quarks
$(\sigma(pp\rightarrow h\chi\chi) \times \text{BR}(h\rightarrow
b\overline{b}))$ are set at 95\% confidence level using the $CL_s$
modified frequentist formalism~\cite{Read:2002hq} with the
profile-likelihood-ratio test statistic~\cite{Cowan:2010js}.  For
the \Zprime-2HDM model, these limits range from 191.3 fb for a
$\Zprime$ mass of 600 GeV and an $A$ mass of 300 GeV to 6.72 fb for a
$\Zprime$ mass of 1600 GeV and an $A$ mass of 600 GeV.  For the vector
mediator model interpretation, the limits range from 1.01 pb for a
mediator mass of 50 GeV and a dark matter mass of 1 GeV to 40.3 fb for
a mediator mass of 800 GeV and a dark matter mass of 500 GeV.  These
are further interpreted as lower limits on the mass parameters of
interest in the specific model. In Figure~\ref{fig:Zprimeplot}
the \Zprime-2HDM exclusion contour in the $(m_{\Zprime},m_{A})$ plane
for $\tan{\beta} = 1$, $m_{\chi}=100\,\gev$ is presented, with limits
more stringent than obtained in Run 1, excluding $Z'$ masses up to
1950 GeV and $A$ masses up to 500 GeV. In Figure~\ref{fig:vectorplot},
the exclusion contour is shown in the $(m_{Z'},m_\chi)$ plane for the
vector mediator model described in Section~\ref{sec:data-mc}.  This
interpretation was not performed in Run 1 and the mass reach for this
choice of couplings excludes $Z'$ masses below 700 GeV for low DM
mass.

%% file: sec-conclusion.tex
A search is presented for dark-matter pair production in association
with a Higgs boson decaying into two $b$-quarks, using 3.2 \ifb\ of
$pp$ collisions collected at $\sqrt{s} = 13$~\TeV~by the ATLAS
detector at the LHC. Two regions are considered, a low-$\Etmiss$
region where the two $b$-quark jets from the Higgs boson decay are
reconstructed separately and a high-$\Etmiss$ region where they are
reconstructed inside a single large-radius trimmed jet.

The data are found to be consistent with the background expectation
and the results are interpreted for two simplified models involving a
massive vector mediator. In the \Zprime-two-Higgs-doublet, constraints
are placed on the $(m_{\Zprime},m_{A})$ space and found to exclude a
wide range of \Zprime\ masses with the pseudo-scalar Higgs mass
exclusion reaching up to 500 GeV. In the context of the vector
mediator model, constraints are placed in the two-dimensional space of
$(m_{Z'},m_{\chi})$ and found to exclude vector mediators with masses
up to 700~GeV.

%% file: Acknowledgements.tex

We thank CERN for the very successful operation of the LHC, as well as the
support staff from our institutions without whom ATLAS could not be
operated efficiently.

We acknowledge the support of ANPCyT, Argentina; YerPhI, Armenia; ARC, Australia; BMWFW and FWF, Austria; ANAS, Azerbaijan; SSTC, Belarus; CNPq and FAPESP, Brazil; NSERC, NRC and CFI, Canada; CERN; CONICYT, Chile; CAS, MOST and NSFC, China; COLCIENCIAS, Colombia; MSMT CR, MPO CR and VSC CR, Czech Republic; DNRF and DNSRC, Denmark; IN2P3-CNRS, CEA-DSM/IRFU, France; GNSF, Georgia; BMBF, HGF, and MPG, Germany; GSRT, Greece; RGC, Hong Kong SAR, China; ISF, I-CORE and Benoziyo Center, Israel; INFN, Italy; MEXT and JSPS, Japan; CNRST, Morocco; FOM and NWO, Netherlands; RCN, Norway; MNiSW and NCN, Poland; FCT, Portugal; MNE/IFA, Romania; MES of Russia and NRC KI, Russian Federation; JINR; MESTD, Serbia; MSSR, Slovakia; ARRS and MIZ\v{S}, Slovenia; DST/NRF, South Africa; MINECO, Spain; SRC and Wallenberg Foundation, Sweden; SERI, SNSF and Cantons of Bern and Geneva, Switzerland; MOST, Taiwan; TAEK, Turkey; STFC, United Kingdom; DOE and NSF, United States of America. In addition, individual groups and members have received support from BCKDF, the Canada Council, CANARIE, CRC, Compute Canada, FQRNT, and the Ontario Innovation Trust, Canada; EPLANET, ERC, FP7, Horizon 2020 and Marie Sk{\l}odowska-Curie Actions, European Union; Investissements d'Avenir Labex and Idex, ANR, R{\'e}gion Auvergne and Fondation Partager le Savoir, France; DFG and AvH Foundation, Germany; Herakleitos, Thales and Aristeia programmes co-financed by EU-ESF and the Greek NSRF; BSF, GIF and Minerva, Israel; BRF, Norway; Generalitat de Catalunya, Generalitat Valenciana, Spain; the Royal Society and Leverhulme Trust, United Kingdom.

The crucial computing support from all WLCG partners is acknowledged gratefully, in particular from CERN, the ATLAS Tier-1 facilities at TRIUMF (Canada), NDGF (Denmark, Norway, Sweden), CC-IN2P3 (France), KIT/GridKA (Germany), INFN-CNAF (Italy), NL-T1 (Netherlands), PIC (Spain), ASGC (Taiwan), RAL (UK) and BNL (USA), the Tier-2 facilities worldwide and large non-WLCG resource providers. Major contributors of computing resources are listed in Ref.~\cite{ATL-GEN-PUB-2016-002}.

%% file: atlas_authlist.tex
\begin{flushleft}
{\Large The ATLAS Collaboration}

\bigskip

M.~Aaboud$^\textrm{\scriptsize 136d}$,
G.~Aad$^\textrm{\scriptsize 87}$,
B.~Abbott$^\textrm{\scriptsize 114}$,
J.~Abdallah$^\textrm{\scriptsize 65}$,
O.~Abdinov$^\textrm{\scriptsize 12}$,
B.~Abeloos$^\textrm{\scriptsize 118}$,
R.~Aben$^\textrm{\scriptsize 108}$,
O.S.~AbouZeid$^\textrm{\scriptsize 138}$,
N.L.~Abraham$^\textrm{\scriptsize 152}$,
H.~Abramowicz$^\textrm{\scriptsize 156}$
% ...
\bigskip
\\
% ...
$^{12}$ Institute of Physics, Azerbaijan Academy of Sciences, Baku, Azerbaijan\\
% ...
$^{65}$ University of Iowa, Iowa City IA, United States of America\\
% ...
$^{87}$ CPPM, Aix-Marseille Universit{\'e} and CNRS/IN2P3, Marseille, France\\
% ...
$^{108}$ Nikhef National Institute for Subatomic Physics and University of Amsterdam, Amsterdam, Netherlands\\
% ...
$^{114}$ Homer L. Dodge Department of Physics and Astronomy, University of Oklahoma, Norman OK, United States of America\\
% ...
$^{118}$ LAL, Univ. Paris-Sud, CNRS/IN2P3, Universit{\'e} Paris-Saclay, Orsay, France\\
% ...
$^{136}$ $^{(a)}$ Facult{\'e} des Sciences Ain Chock, R{\'e}seau Universitaire de Physique des Hautes Energies - Universit{\'e} Hassan II, Casablanca; $^{(b)}$ Centre National de l'Energie des Sciences Techniques Nucleaires, Rabat; $^{(c)}$ Facult{\'e} des Sciences Semlalia, Universit{\'e} Cadi Ayyad, LPHEA-Marrakech; $^{(d)}$ Facult{\'e} des Sciences, Universit{\'e} Mohamed Premier and LPTPM, Oujda; $^{(e)}$ Facult{\'e} des sciences, Universit{\'e} Mohammed V, Rabat, Morocco\\
% ...
$^{138}$ Santa Cruz Institute for Particle Physics, University of California Santa Cruz, Santa Cruz CA, United States of America\\
% ...
$^{152}$ Department of Physics and Astronomy, University of Sussex, Brighton, United Kingdom\\
% ...
$^{156}$ Raymond and Beverly Sackler School of Physics and Astronomy, Tel Aviv University, Tel Aviv, Israel\\
% ...
$^{a}$ Also at Department of Physics, King's College London, London, United Kingdom\\
$^{b}$ Also at Institute of Physics, Azerbaijan Academy of Sciences, Baku, Azerbaijan\\
$^{c}$ Also at Novosibirsk State University, Novosibirsk, Russia\\
$^{d}$ Also at TRIUMF, Vancouver BC, Canada\\
$^{e}$ Also at Department of Physics {\&} Astronomy, University of Louisville, Louisville, KY, United States of America\\
% ...
\end{flushleft}
